\documentclass[aps,prl,reprint,amsmath,nofootinbib,superscriptaddress,showkeys]{revtex4-1}
\usepackage[utf8]{inputenc}
\usepackage{indentfirst}
\usepackage{amssymb}
\usepackage{amsmath,amsthm,latexsym,bm}
\usepackage{graphicx}
\usepackage{float}
\usepackage[margin=1.0in]{geometry}
\usepackage{bbold,yfonts}
\usepackage[dvipsnames]{xcolor}
\usepackage{tabularx}
\usepackage{slashed}
\usepackage{upgreek}
\usepackage{cancel}
\usepackage{bigints}
\usepackage{placeins}
\usepackage[normalem]{ulem}
\usepackage{geometry}
\usepackage{extarrows}
\usepackage[colorlinks=true,backref=false, linktocpage=true,
citecolor=blue,urlcolor=blue,linkcolor=blue,pdfpagemode=UseOutlines,pdfstartview=FitH,bookmarksopen]{hyperref}
\newcommand\sect[1]{\noindent \textbf{\emph{#1.}}}

\begin{document}

\title{Phase transitions and finite-size effects in integrable virial statistical models}

\author{Xin An}
\email{xin.an@ugent.be}
\affiliation{Department of Physics and Astronomy, Ghent University, 9000 Ghent, Belgium}
\affiliation{National Centre for Nuclear Research, 02-093 Warsaw, Poland}

\author{Francesco Giglio}
\email{francesco.giglio@glasgow.ac.uk} 
\affiliation{School of Mathematics and Statistics, University of Glasgow, Glasgow, United Kingdom}

\author{Giulio Landolfi}
\email{giulio.landolfi@le.infn.it, giulio.landolfi@unisalento.it}
\affiliation{Dipartimento di Matematica e Fisica
``Ennio De Giorgi" Universit\a`a del Salento \\
and I.N.F.N. Sezione di Lecce, via Arnesano I-73100 Lecce, Italy}

\date{\today}

\begin{abstract}
\noindent
We analyze thermodynamic models for fluid systems in equilibrium based on a virial expansion of the internal energy in terms of the volume density. We prove that the models, formulated for finite-size systems with $N$ particles, are exactly solvable to any expansion order, as expectation values of physical observables (e.g., volume density) are determined from solutions to nonlinear {\em C-integrable} partial differential equations (PDEs) of hydrodynamic type.  In the limit $N\to \infty$, phase transitions emerge as classical shock waves in the space of thermodynamic variables. Near critical points, we argue that the volume density exhibits a scaling behavior consistent with the {\em Universality Conjecture} in viscous transport PDEs. As an application, we employ our framework to nuclear and quark matter, constructing a global quantum chromodynamics (QCD) phase diagram that reveals critical points for the nuclear liquid-gas transition and the hadron gas–quark-gluon plasma transition. We demonstrate how finite-size effects smear critical signatures, implying their potential impact on the search for the QCD critical point.
\end{abstract}

\maketitle

\sect{Introduction}
Understanding phase transitions in matter is a fundamental subject in physics. The presence of interparticle interactions, as well as the involvement of several thermodynamic functions, can result in effects at different scales, with possible
multiple critical points and distinct ground states. Variation of a thermodynamic function, such as temperature or pressure, can suppress one phase while promoting another, unveiling a rich interplay between competing phases~\cite{ParisiRevModPhys.95.030501,Zinn-Justin:2002ecy,PELISSETTO2002549}. Such phenomena are ubiquitous in semiconductors~\cite{PhysRevLett.44.182}, crystal field~\cite{SumedhaPRE2020}, spin systems~\cite{PhysRevLett.69.221},
polymorphic fluids~\cite{LLScience2020}, and matter in quantum chromodynamics (QCD)~\cite{PhysRevC.82.055205}. Despite the numerous advances in extensive theoretical and experimental studies of these systems, there remains however a need for viable phenomenological approaches and global exact results beyond the regime of the critical point and the thermodynamic limit (TL), which largely rely on costly lattice simulations~\cite{Nagata:2021ugx}. 
 
In this Letter, we analyze the thermodynamics of fluid systems where the nontrivial interactions among particles result in an effective potential energy in the form of a virial-like expansion in the order parameter. 
Our approach, inspired by papers~\cite{MORO2014, BARRA2015290}, will enable us to identify partial differential equations (PDEs) for the associated statistical models, and to predict behavior of their solutions globally and locally even beyond TL, when finite-size effects are foreseen~\cite{YangLee_PhysRev.87.404,
Borrmann_PhysRevLett.84.3511,Bernhardt:2021iql, Kovacs:2023kbv,LorenzoniPhysRevE.100.022103}. Such effective statistical models provide a simple yet powerful description for van der Waals (vdW) or more {\em imperfect} fluid systems across various areas of physics, including classical liquid-gas (L-G) systems~\cite{callen}, the rotating or charged anti-de Sitter black holes~\cite{Chamblin:1999tk}, astrophysical objects~\cite{Oertel}, ultracold quantum gases~\cite{Chin_2010} and last but not least, dense nuclear and quark matter in QCD~\cite{Kapusta:2006pm} where the finite-size effects are expected to be significant---the major motivation for this work.

\sect{Virial-like statistical models}
Phase transitions are conveniently studied within the framework of statistical mechanics following the maximum entropy principle~\cite{callen}. 
This principle turns into an extremity problem for the proper entropic potential $\Psi$ for a specific statistical ensemble, which is related to the system's entropy $S$---the relevant potential in the {\em microcanonical ensemble} whose natural variables are a set of conserved quantities denoted by vector $\Lambda$---by Legendre transform $\Psi(\tau)=\sup_{\Lambda}\,[-S(\Lambda) + \tau\Lambda]$, where $\tau\Lambda=\sum_{j=1}^{J}\tau_j\Lambda_j$ is a product of $J$ conjugate pairs $\{\Lambda_j, \tau_j \}$. The probability of any available microscopic state labeled by $i$ is weighed as $p_i=e^{-\tau\Lambda^{(i)}}/{\mathcal Z}$, 
normalized by the partition function associated with the potential $\Psi(\tau)$ to ensure $\sum_i p_i=1$:
\begin{equation*}\label{eq:Z}
    \mathcal Z(\tau)=e^{-\Psi(\tau)}=\sum_{i} e^{-\tau\Lambda^{(i)}}\propto\int d\Lambda\,e^{S(\Lambda)-\tau\Lambda}\,,
 \end{equation*}
where in the last step we have taken the continuum limit of the summation over all microscopic states to the integration over $d\Lambda=d\Lambda_1\dots d\Lambda_{J}$ and dropped a multiplicative constant. Once the partition function for a given system is assigned, all thermodynamic properties follow~\cite{lebellac}.

For a generic hard spheres fluid system, the microcanonical entropy (up to irrelevant constants) has the well-known form
\begin{equation}\label{eq:S_mc}
    S(E,V,N)=N\ln\left[\frac{V-Nb}{N}\left(\frac{E-U(V,N)}{N}\right)^{3/2}\right]\,,
\end{equation}
which depends on its natural (extensive) variables such as internal energy $E$, volume $V$, and particle number $N$. The parameter $b$ identifies the minimum volume density\footnote{Throughout this work by \emph{density} we mean \emph{per particle}.} of the configurations in the ensemble. The total potential energy $U$  is assumed a virial-type form 
\begin{equation}\label{eq:U(V)}
    U(V,N)=-\sum_{j=1}^{m} \frac{N^{j+1}a_{j+1}}{j V^j}\, , 
\end{equation}
where the virial expansion coefficients $a_j$, for $j=1,\dots, m$, are real constants, since for hard core (or short range) potentials, these coefficients do not (approximately) depend on temperature.\footnote{Virial coefficients may acquire temperature dependence via, e.g., corrections from quantum statistics \cite{Vovchenko:2015vxa}, which is beyond the scope of this work.} The case $m=1$ reproduces the potential energy of the celebrated vdW model. From a mean-field perspective, the potential $U$ can be seen to account for the dependence of the averaged particles interaction upon their average distance $\bar{r} \sim v^{1/3}$, where $v\equiv V/N$ denotes the volume density.

For the case considered in this Letter, where the energy $E$ and volume $V$ of the fluids vary while the particle number $N$ is kept fixed, it is more convenient to consider the {\em isothermal-isobaric ensemble} where its natural (intensive) variables, $t=(\partial_E S)_{V}$ and $x=(\partial_V S)_{E}$, control the fluctuations of energy and volume respectively.\footnote{For a one-component fluid system, two independent intensive degrees of freedom are needed according to \emph{Gibbs phase rule}, implying $J=2$.} In terms of standard pressure $P$ and temperature $T$, $t=1/T$ and $x=P/T$. This choice has been shown to arise naturally in the treatment of vdW~\cite{MORO2014, BARRA2015290} and nematic fluids~\cite{DEMATTEIS2018}. Thus we are led to consider the scaled Gibbs free energy (i.e., Gibbs free energy divided by temperature)
\begin{equation*}\label{eq:GN}
    \mathcal G_N(t,x)\equiv\mathcal G(t,x;N)=\sup_{E,V}\,[-S(E,V,N)+tE+xV]\,,
\end{equation*}
from which the partition function is defined by\footnote{The boundary effects in finite systems can be described by introducing in the free energy a surface energy term suppressed by $N^{-1/3}$~\cite{Bugaev:2000ar,Lee:2000hj}.}
\begin{equation}\label{eq:Z_N_general}
    \mathcal Z_N(t,x)\equiv e^{-\mathcal G_N(t,x)}=\int dEdV \, e^{S(E,V,N)-tE-xV}\,.
\end{equation}
 Since $U=U(V,N)$, the integration over $E$ accounts only for the kinetic part $K=E-U$. This contributes an irrelevant normalization factor $c(t)
 \propto t^{-3N/2}$ to the partition function~\eqref{eq:Z_N_general}, provided that the observables $O$ under consideration are independent of $K$. Consequently, the integration in Eq.~\eqref{eq:Z_N_general} reduces to one over all ensemble configurations labeled by $V$ (or $v$ for fixed $N$) alone and one can define another partition function $Z_N$ via $\mathcal{ Z}_N=c(t)Z_N$:
\begin{multline}\label{eq:Z_N}
    Z_N(t,x)\equiv e^{-G_N(t,x)}\equiv e^{-Ng_N(t,x)}\\
    =\int dv \, e^{-N\psi(t,x;v)}\,,
\end{multline}
where $G_N$ and $g_N$ denote the associated free energy and its density respectively, while 
\begin{equation}\label{eq:psi(v)}
    \psi(t,x;v)=-s(v)+t\epsilon(v)+xv
\end{equation}
is a potential density function. The {\em expectation value} of observable
$O_N(v)=O(V,N)$ is thus given by\footnote{Generically, $O=O(E,V,N)$. For example, the average kinetic energy $\langle{K(E,V,N)\rangle}\equiv\langle{E-U(V,N)\rangle}=3N/2t$, being consistent with the equipartition theorem.}
\begin{equation}\label{eq:<O>}
    \langle O_N\rangle= \frac{1}{Z_N} \int dv\, e^{-N\psi(t,x; v)} O_N(v) \,.
\end{equation}
Accordingly, we focus on the configurational entropy and potential energy density given by
\begin{equation}\label{eq:s_e}
    s(v)=\ln(v-b)\,, \qquad \epsilon(v)=-\sum_{j=1}^{m} \frac{a_{j+1}}{j v^j}\,,
\end{equation}
which result from Eqs.~\eqref{eq:S_mc} and \eqref{eq:U(V)} respectively, and depend only on $v$.

One key observation of our work is that the partition function \eqref{eq:Z_N} fulfills the  $(m+1)-$th order linear PDE: 
\begin{equation}\label{eq:pde_Z_N}     
    \left(\partial_x^{m}\partial_t+\sum_{j=1}^m\frac{(-N)^{j+1}a_{j+1}}{j}\partial_x^{m-j}\right)Z_N(t,x)=0\,.
\end{equation}
Eq.~\eqref{eq:pde_Z_N} is singular in TL $N\to\infty$. It is therefore convenient to identify the differential problem for the free energy density $g_N$ via Eq.~\eqref{eq:Z_N}.  Direct computations give the following nonlinear PDE: 
\begin{multline}\label{eq:pde_g_N}
    B_m \partial_t g_N \\
     +\,\sum_{j=1}^m
    B_{m-j} \, \left[\binom{m}{j}\partial_x^j\partial_tg_N+\frac{(-N)^{j}a_{j+1}}{j}\right]=0\,,
\end{multline}
where $B_n=n! \sum_{\{\lambda^{(n)}_{\bf j} \} } \,\prod_{i=1}^{n}  \frac{1}{(i!)^{j_i} j_i !}  \left(-N\partial^i_x g_N \right)^{j_i}$
is the $n$th complete exponential Bell polynomial and the summation is performed over all index partitions $\lambda^{(n)}_{\bf j}=\{j_1, \dots,j_n\}$ satisfying $\sum_{i=1}^n i j_i=n$. When $n=0$, one has, by definition, $B_0=1$. Using Eqs.~\eqref{eq:Z_N}, \eqref{eq:psi(v)} and \eqref{eq:<O>}, the averaged volume density and energy density are simply obtained via the derivative of $g_N$:
\begin{subequations}\label{eq:v_N_e_N}
    \begin{align}
     {\rm v}_N\equiv\langle v_N\rangle=\partial_xg_N\,, \label{eq:v_N}\\
     \varepsilon_N\equiv\langle \epsilon_N\rangle=\partial_tg_N\,,
\end{align}
\end{subequations}
which allows us to convert Eq.~\eqref{eq:pde_g_N} into a PDE for ${\rm v}_N$. It follows immediately from Eqs.~\eqref{eq:v_N_e_N} that $\partial_t {\rm v}_N=\partial_x \varepsilon_N$. 

We emphasize that the PDE \eqref{eq:pde_Z_N} for the partition function is linear for the whole family of statistical models admitting the arbitrary virial-like expansion \eqref{eq:U(V)}. Since physical observables connect to the partition function by transformations involving variable changes and derivatives such as those in Eqs.~\eqref{eq:v_N_e_N}, the whole family of statistical models under consideration is therefore {\em integrable} (more precisely {\em C-integrable}~\cite{Calogero1991WhyAC}). In particular, the volume density at finite $N$ is immediately obtained from the partition function via [cf. Eq.~\eqref{eq:v_N}]
\begin{equation*}\label{eq:Cole-Hopf}
  {\rm v}_N=-\frac{1}{N} \partial_x \ln{Z_N}\,,
\end{equation*}
known as Cole--Hopf-type transformation in the language of integrable systems~\cite{Calogero1991WhyAC}.

\sect{Multiple phases at large $N$}
Letting $g_N\to g, {\rm v}_N\to {\rm v}, \varepsilon_N\to\varepsilon$ in TL, and inserting it into Eq.~\eqref{eq:pde_g_N}, one obtains the identity $\varepsilon=\partial_t g=\epsilon(\partial_x g)=\epsilon ({\rm v})$, where Eq.~\eqref{eq:s_e} is used. Applying Eqs.~\eqref{eq:v_N_e_N}, one arrives at the following local conservation law:
\begin{equation}\label{eq:pde_v}
     \partial_t{\rm v}-\epsilon'({\rm v}) \, \partial_x{\rm v}=0\,.
\end{equation}
This is the Maxwell closure relation for the Gibbs free energy in TL. Its general solution can be obtained via the method of characteristics, yielding $x+ t \epsilon'({\rm v})-f({\rm v})=0$ where $f({\rm v})$ is an arbitrary function that can be specified by assigning the Cauchy condition, as for instance, if a particular isothermal curve is known in some thermodynamic regimes [such as possibly known isotherms $x=f({\rm v})$ in the regime of very high temperature]. Notice that the function $f$ relates to the function $s$ in Eq.~\eqref{eq:psi(v)} as $f({\rm v})=s'({\rm v})$. 
Indeed, by employing the \emph{Laplace's formula} to $Z_N$, we get 
\begin{equation}\label{eq:Z_N_SPA}
    Z_N(t,x)\simeq \sum_\ell \sqrt{\frac{2\pi}{N \psi''_\ell}}e^{-N\psi_\ell}\,,
\end{equation}
where the sum runs over the local minima of the potential $\psi$ defined in Eq.~\eqref{eq:Z_N}, $\psi_{\ell}\equiv\psi(t,x;{\rm v}_\ell)$, with $\ell$ labeling the stationary points ${\rm v}_\ell$ at which
\begin{equation}\label{eq:eos_SPA}
    C({\rm v}; t,x)\equiv  \partial_{\rm v}\psi=x+t \epsilon'({\rm v})-s'({\rm v})=0\,.
\end{equation}  
Evidently, Eq.~\eqref{eq:eos_SPA} is just the implicit solution to Eq.~\eqref{eq:pde_v} upon identification $f({\rm v})=s'({\rm v})$. The implicit solutions given by  $C({\rm v})=0$ represent changes of the volume density of the nonlinear-wave type, which generically undergo a gradient catastrophe~\cite{whitham,Thom} that emerges at the inflection points (i.e., thermodynamic critical point)~\cite{MORO2014}, identified by the conditions 
\begin{equation}\label{eq:inflection_points}
    C({\rm v})= C' ({\rm v})=C'' ({\rm v})=0\,.
\end{equation}
As discussed in \cite{MORO2014,BARRA2015290,Giglio2016IntegrableEV}, thermodynamic critical points represent the onset of classical (viscous)
shock waves~\cite{whitham} where order parameters and/or their derivatives experience jumps, hence unveiling the occurrence of phase transitions.
However, when multiple solutions exist, the equation of state (EOS) does not contain information on which one is the most likely to occur. From a nonlinear wave viewpoint, this problem is equivalent to that of shock fitting, i.e., replacing a multi-valued solution with a discontinuous one (a shock wave), with the discontinuity located according to some specific entropic criteria (cf. Maxwell construction)~\cite{Bressan}. 
 
In TL, multiple solutions of EOS~\eqref{eq:eos_SPA}, associated with different thermodynamic phases, may exist. A criterion for identifying the favored solution discerns from Eq.~\eqref{eq:Z_N_SPA}, which for very large $N$ yields 
\begin{equation}\label{eq:v_N_SPA}
    {\rm v}_N(t,x)\simeq\frac{\sum_k {\rm v}_k(t,x) (\psi''_k)^{-1/2}\, e^{-N \psi_k}}{\sum_\ell (\psi''_\ell)^{-1/2}\,e^{-N \psi_\ell}}\simeq \bar{\rm v}_m(t,x)\,,
\end{equation}
where $\bar{\rm v}_m$ is the solution to $\psi'=0$ that corresponds to the lowest minimum of potential $\psi({\rm v})$. Therefore,  the phase diagram is identified by selecting the global minimum for each pair of $t$, and will display coexistence curves when two minima are resonant, that is when $\psi_j=\psi_k$ for two distinct ${\rm v}_j$ and ${\rm v}_k$. 
Reconstructing coexistence lines $\mathcal{L}_{jk}\equiv\{(t,x)\in\mathbb{R}^+\times \mathbb{R}\,|\,\psi'({\rm v}_j)=\psi'({\rm v}_k)=0,\,\,\psi({\rm v}_j)=\psi({\rm v}_k)\}$ for pairs $j\neq k$, is mathematically equivalent to  the problem of tracking  trajectories of the underlying shock wave~\cite{MORO2014,Giglio2016IntegrableEV,callen}. When $L$ equally deep lowest minima exist, Eq.~\eqref{eq:v_N_SPA} implies that all of them contribute to the large $N$ asymptotic solution for the volume density, with those exhibiting the flattest convex shapes in $\psi$ contributing the most.
Indeed, from Eq.~\eqref{eq:v_N_SPA}, we get ${\rm v}_N\simeq \sum_{\ell=1}^L \gamma_\ell {\rm v}_\ell$ as $N\to\infty$, where the $\gamma_k$'s are weights defined by $\gamma_k=(\psi''_k)^{-1/2}/\sum_{\ell=1}^L (\psi''_\ell)^{-1/2}$. Being single valued and continuous, this asymptotic solution gives a volume density that lies on the isothermal curve, representing an actual state of the system, differently from the case for multiphase systems~\cite{callen}.

\sect{Universality near the critical points}
In the large but finite $N$ limit, following Eqs.~\eqref{eq:pde_g_N} and~\eqref{eq:v_N_e_N}, one finds that the leading correction to the equation for volume density ${\rm v}_N$ can be encoded into a $1/N$ viscous term [cf. Eq.~\eqref{eq:pde_v}]:
\begin{equation}\label{eq:pde_v_visc}
    \partial_t {\rm v}_N -\epsilon'({\rm v}_N) \partial_x {\rm v}_N  =
-\frac{1}{2N}\partial_x\left[\epsilon''({\rm v}_N)\partial_x {\rm v}_N \right] \,  
+h.o.t.,        
\end{equation} 
where the higher order [$\mathcal O(1/N^k)$ with $k\geq 2$] terms are given by functions of ${\rm v}_N$ and their derivatives.\footnote{
The PDE for ${\rm v}_N$ arising from \eqref{eq:pde_g_N} contains 
mixed derivatives of ${\rm v}_N$ with respect to $t$ and $x$. These can be approximated by exploiting the equation itself, which has the structure $\partial_t {\rm v}_N\simeq \epsilon'({\rm v}_N)\partial_x{\rm v}_N+O(1/N)$, thus leading to Eq.~\eqref{eq:pde_v_visc}.}

\begin{figure}[!t]
\centering
    \includegraphics[width=0.237\textwidth]{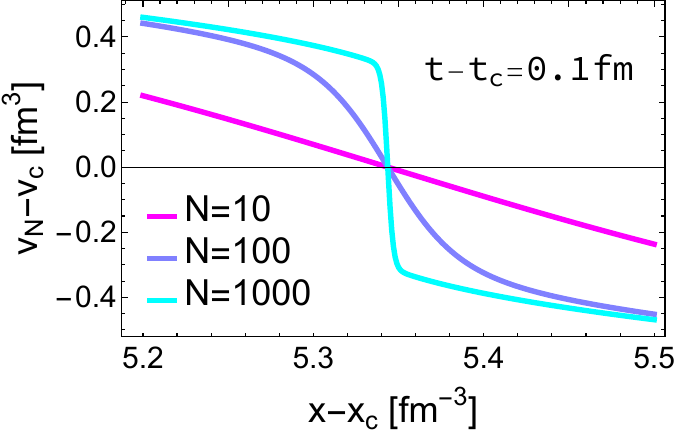}
    \includegraphics[width=0.237\textwidth]{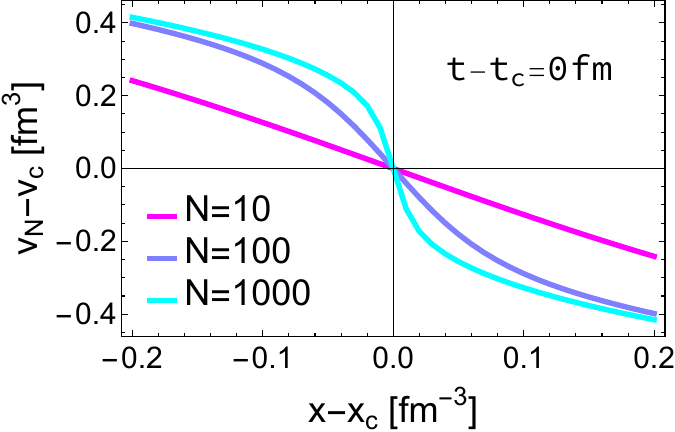}
\caption{Universal behavior and finite-size corrections in the vicinity of QCD critical points at $t-t_c=0.1$ fm (left) and $t-t_c=0$ fm (right). The parameters are chosen from Eq.~\eqref{eq:coeffs} for the QCD critical point.}
\label{fig:plot_v_finite_N}
\end{figure}

Finding the generalized Burgers equation \eqref{eq:pde_v_visc} enables to argue on the expected modifications to the fluid EOS when $N$ is finite in terms of a well-posed mathematical problem. In particular, it is well understood that the $1/N$ viscous term in Eq.~\eqref{eq:pde_v_visc} prevents the appearance of singular behavior at the critical points that would originate in the ``inviscid limit'' at $N\to \infty$. In addition, fundamental results known in the PDE's literature can be adapted to describe the behavior as one approaches the critical region. More precisely, a connection is established with a remarkable mathematical result known in the theory of integrable systems as {\em Universality Conjecture}~\cite{dubrovin2006, dubrovin2012}. 
Following~\cite{dubrovin2012}, we are therefore led to deduce that near a given critical point located at $t_c, x_c,{\rm v}_c$, the solution to Eq.~\eqref{eq:pde_v_visc} behaves as
\begin{multline}\label{eq:critical_v_DE}
   {\rm v}_N(t,x)={\rm v}_c + \frac{\gamma}{N^{1/4}} \, 
        U\left( \frac{\Delta x +\epsilon'({\rm v}_c)\Delta t}{\alpha N^{-3/4}}, 
         \frac{\Delta t}{\beta N^{-1/4}}\right) \\
         + O(N^{-1/2})\,,         
\end{multline}
where $\Delta x=x-x_c$, $\Delta t=t-t_c$, and the function $U(r_1,r_2)$ is defined via logarithmic derivative of the Pearcey function: 
\begin{equation*}\label{eq:U(r1,r2)}
    U(r_1,r_2)=-2 \partial_{r_1} \ln \int^\infty_{-\infty} dr e^{ -\frac{1}{8}(r^4-2r^2 r_2+4r r_1)}\,. 
\end{equation*}
The scale parameters in Eq.~\eqref{eq:critical_v_DE} are specified via $\alpha=\frac{1}{2\gamma}=(\frac{\kappa}{8})^{1/4}, \beta=(\frac{\kappa}{2[\epsilon''({\rm v}_c)]^2})^{1/2}$, where $\kappa=\frac{1}{6} \left[\epsilon^{(4)}({\rm v}_c) \, t_c - s^{(4)}({\rm v}_c)\right]$. 
The ``critical EOS'' \eqref{eq:critical_v_DE} at various values of $N$ is shown in Fig.~\ref{fig:plot_v_finite_N}, illustrating how the critical signature smooths out at small $N$~\cite{Chomaz:1999az}.

\sect{Application: phase transitions in QCD}
An important example of a physical system displaying multiple critical points is represented by nuclear matter whose underlying physics is described by QCD~\cite{Stephanov:2006zvm}. The two major properties of quarks and gluons, i.e., asymptotic freedom at high energy and confinement at low energy, indicate two different phases separated by a first-order phase transition: the quark-gluon plasma (QGP) and hadron gas (HG)~\cite{Sorensen:2023zkk}. The latter also features a vdW-type nuclear L-G transition at low energy due to hadronic interactions~\cite{Panagiotou:1984rb,PhysRevLett.75.1040,Wang:2020tgb}. Effective mean-field models for describing both critical points are feasible by choosing baryon density $n_B$ as the order parameter which is well defined across different scales. Indeed, it has been emphasized that mean-field statistical mechanics approaches to QCD matter prove to be very effective because they naturally lead to EOS reproducing the known properties of ordinary nuclear matter, and, as an effective and macroscopic statistical model, allow us to explore critical behavior in dense nuclear matter over vast regions of the QCD phase diagram~\cite{SorensenPhysRevC.104.034904, PhysRevLett.131.202303}.

\setlength{\tabcolsep}{5pt}
\renewcommand{\arraystretch}{1.2}
\begin{table}[b]
\centering
    \begin{tabular}{c | c | c | c} 
     \hline\hline
    ~Quantity~ & ~Value~ & ~Quantity~ & ~Value~ \\
     \hline
    $T_{c,1}$ & 0.10 fm$^{-1}$ & $T_{c,2}$ & 0.51 fm$^{-1}$ \\
     \hline
    ${\rm v}_{c,1}$ & 16.45 fm$^{3}$ & ${\rm v}_{c,2}$ & 2.08 fm$^{3}$ \\ 
     \hline
    ${\rm v}_a$ & 2.40 fm$^{3}$ & ${\rm v}_b$ & 1.89 fm$^{3}$\\
     \hline\hline
    \end{tabular}
\caption{Physical values chosen to determine coefficients in Eq.~\eqref{eq:coeffs}.}
\label{tab:conditions}
\end{table}

\begin{figure}[!b]
\centering{
\includegraphics[width=0.237\textwidth]{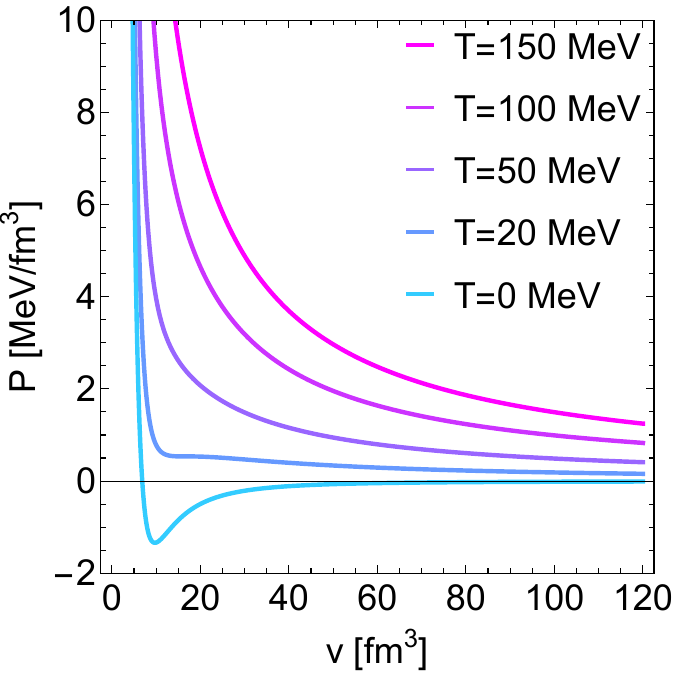}
\includegraphics[width=0.237\textwidth]{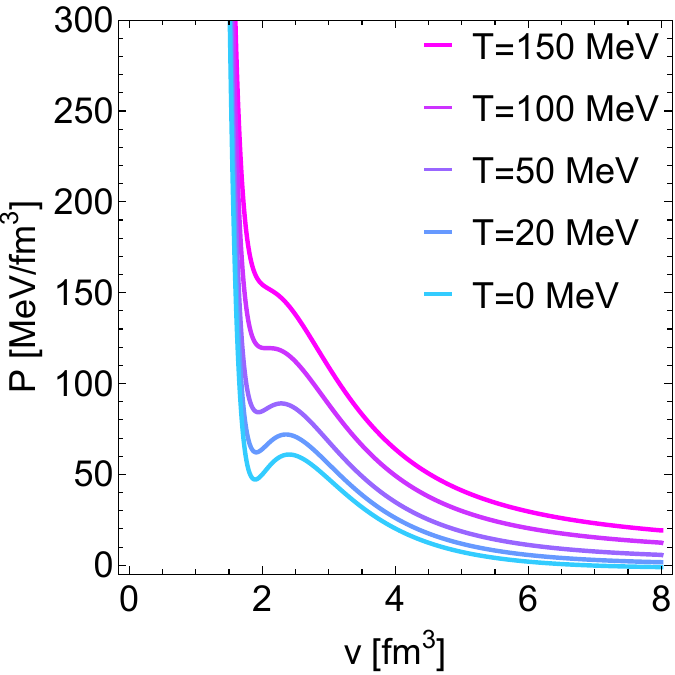}}
\caption{Isothermal curves predicted by the EOS~\eqref{eq:eosQCD} with parameters \eqref{eq:coeffs} at different thermodynamic scales. 
The left and right panels show the nuclear L-G and the HG-QGP transitions, respectively.}
\label{fig:P_v}
\end{figure}

In the isothermal-isobaric ensemble, we identify $N$ as the total baryon number, which remains fixed due to baryon number conservation. Consequently, the order parameter is taken to be the volume density ${\rm v}=1/n_B$. We therefore propose a phenomenological mean-field model by assuming that the phase transitions governed by QCD can be described by such order parameter via the hard-sphere entropy and potential energy \eqref{eq:s_e}, taking a virial-type expansion with $m=6$. Accordingly, in TL, Eq.~\eqref{eq:eos_SPA} implies the following EOS 
\begin{equation}\label{eq:eosQCD}
    P({\rm v},T)=\frac{T}{{\rm v}-b}-\sum_{j=2}^6\frac{a_j}{{\rm v}^j}\,.
\end{equation}
To determine the parameters $b$ and $a$s, six independent conditions shall be imposed. First, the application of Eq.~\eqref{eq:inflection_points} requires $\partial_{\rm v} P({\rm v}_{c,l}, T_{c,l})=\partial_{\rm v}^2 P({\rm v}_{c,l}, T_{c,l})=0$ for the nuclear L-G and HG-QGP transition labeled by $l=1,2$ respectively. The critical volume densities are set to ${\rm v}_{c,1}\simeq1/(0.38n_0)$ at $T_{c,1}=20~\text{MeV}\simeq0.10~\text{fm}^{-1}$ and ${\rm v}_{c,2}\simeq1/(3n_0)$ at $T_{c,2}=100~\text{MeV}\simeq0.51~\text{fm}^{-1}$, where $n_0\simeq
0.16/\text{fm}^{3}$ is the nuclear saturation density~\cite{Fukushima_2011}. Furthermore, at $T=0$ MeV,
the two spinodal points~\cite{An:2017brc} for the HG-QGP transition are located at ${\rm v}_a=1/(2.6n_0)$ and ${\rm v}_b=1/(3.3n_0)$, with $\partial_{\rm v} P({\rm v}_a,0)=\partial_{\rm v} P({\rm v}_b,0)=0$. From the above conditions whose explicit values are listed in Table~\ref{tab:conditions}, one obtains 
\begin{align}\label{eq:coeffs}
    a_2\!&\simeq 16{\rm v}_0\,\text{MeV}, \,a_3 \!\simeq 108{\rm v}_0^2\,\text{MeV}, \,a_4 \!\simeq \!\!-213{\rm v}_0^3\,\text{MeV}, \notag\\
    a_5\!&\simeq 93{\rm v}_0^4\,\text{MeV},\,a_6 \!\simeq \!\! -12{\rm v}_0^5\,\text{MeV}, \,
 b\! \simeq 0.075 {\rm v}_0,  
\end{align}
where ${\rm v}_0=1/n_0$.
The resulting isothermal curves of the EOS \eqref{eq:eosQCD}, in the low- and high-volume density regimes, are displayed in Fig.~\ref{fig:P_v}. 

\begin{figure}[!t]
\centering{
\hspace{-1cm}
\includegraphics[width=0.4\textwidth]
{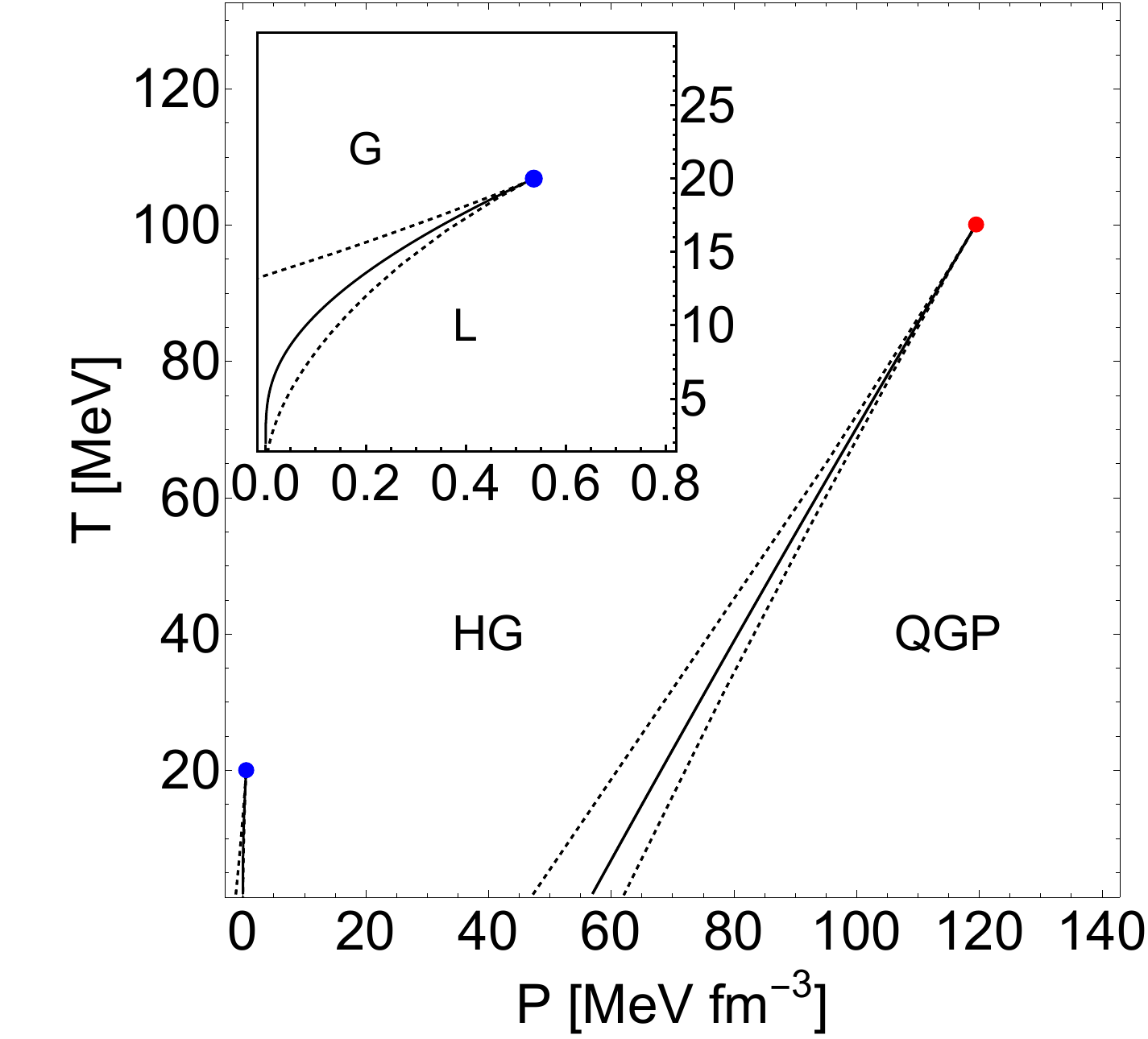}}
\caption{
QCD phase diagram in the $T$-$P$ plane, predicted by EOS~\eqref{eq:eosQCD} with parameters specified in Eq.~\eqref{eq:coeffs}. The blue and red circles represent the nuclear L-G and HG-QGP critical points, respectively, each associated with its corresponding coexistence line (solid) and spinodal lines (dashed).}
\label{fig:phase diagram}
\end{figure}

\vspace{20pt}

\begin{figure}[!t]
\centering{
\includegraphics[width=0.45\textwidth]
{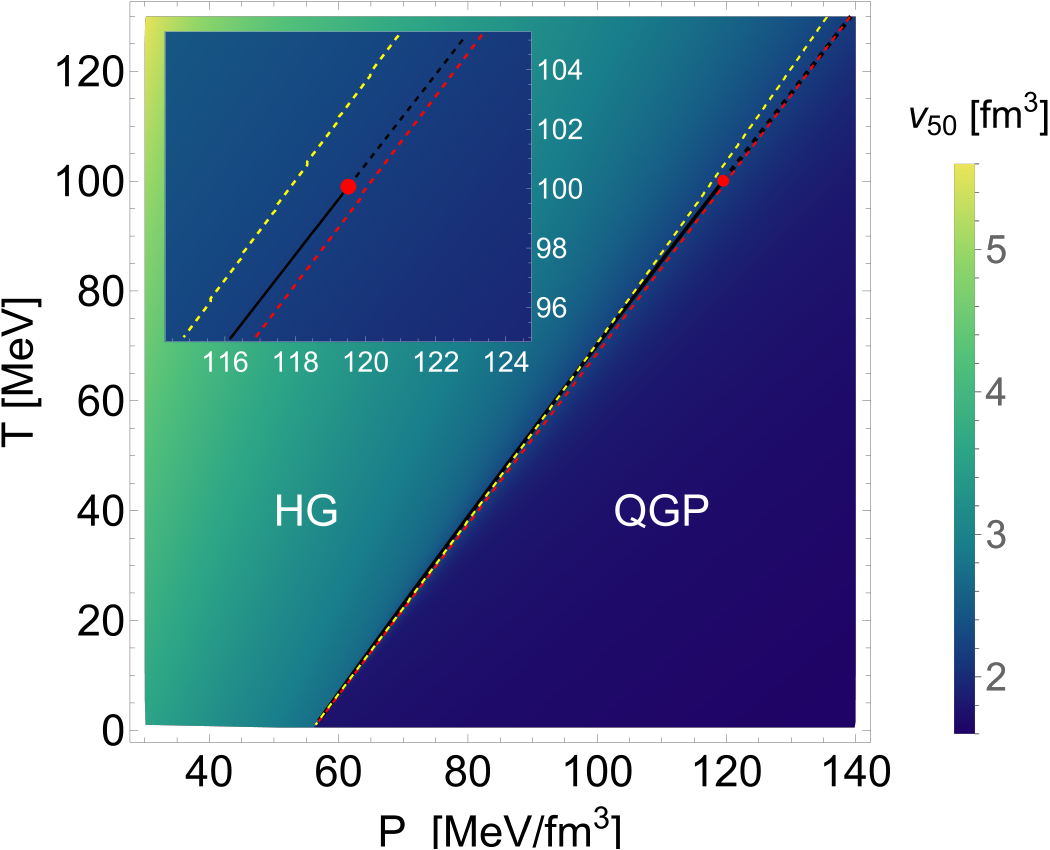}}
\caption{Volume density at finite size 
around the HG-QGP transition region obtained as solution to the differential problem implied by Eqs.~\eqref{eq:v_N_e_N}. Virial parameters are taken the same as in the TL case, Eq.~\eqref{eq:coeffs}. The loci of maxima of specific heat (dashed red) and isothermal compressibility (dashed yellow) at $N=50$, together with the coexistence (solid black) and crossover (dashed black) lines in TL are displayed.}
\label{fig:phase diagram N}
\end{figure}

The QCD phase diagram in the $T$-$P$ plane and its remnant at finite $N$ are shown in Figs.~\ref{fig:phase diagram} and~\ref{fig:phase diagram N}, respectively. While sharp thermodynamic singularities and phase boundaries are well defined in TL, only their remnants, identified by the locations of the maxima of thermodynamic gradients
(e.g., isothermal compressibility or specific heat) persist at finite $N$. As a result, the critical behavior is smoothed out: the singularity at the critical point is smeared, and the TL coexistence curve also turns into a crossover---a remnant of the discontinuity whose location shifts in the phase diagram relative to its position in TL. Our approach extends crossover lines in the TL to finite $N$ and shows that different criteria yield notably distinct predictions when finite-size corrections are taken into account, in contrast to the TL case, where the predictions are indistinguishable near the critical point~\cite{doi:10.1073/pnas.0507870102}. We also observed that the Widom lines, determined by the maxima of response functions (e.g., specific heat and isothermal compressibility), do not merge at the critical point as in the TL. Furthermore, with the chosen parameters, the smaller shift of the Widom line determined by the specific heat suggests it as a more accurate estimate of the crossover in the TL (see Fig.~\ref{fig:phase diagram N})~\cite{PhysRevLett.112.135701}.

\sect{Conclusions}
We have developed a framework for discussing fluid systems whose internal energy can be written in the virial-type form \eqref{eq:U(V)}. In particular, we have determined a linear PDE for the partition function, and {\em C}-integrable PDEs for the free energy and order parameter (volume density in the isothermal-isobaric ensemble). Gradient catastrophe phenomena and
singular solution behaviors, implying discontinuous phase transitions, are allowed in TL $N\to \infty$. Remarkably, phase transitions with discontinuities are instead prevented when finite $N$ leading corrections are retained, and a behavior in agreement with the Universality Conjecture~\eqref{eq:critical_v_DE} is expected in the vicinity of a critical region. Furthermore, our approach leads to prediction of the volume density along the coexistence lines at finite $N$ in terms of the thermodynamic potential around competing phases. 

The formalism has then been implemented into the phase behavior of QCD, which remains poorly understood to date~\cite{Stephanov:2006zvm,Fukushima_2011,Guenther:2020jwe}. Our findings on finite-size effects lead to the conclusion that the search for the QCD critical point in experiments might be even more challenging than figured insofar. Indeed, these effects, varying with the system size, may obscure key characteristic signatures such as the enhanced fluctuations of conserved quantities. 
Consequently, interpreting experimental data, including those from RHIC Beam Energy Scan program~\cite{STAR:2025zdq}, NA61/SHINE experiment~\cite{NA61SHINE:2025whi}, and HADES experiment~\cite{HADES:2020wpc}, requires careful consideration of finite-size effects to ensure reliable conclusions.

Our formalism can be complemented by extending it to ensembles described by different thermodynamic variables, such as chemical potentials and magnetic field, and incorporated into hydrodynamic simulations for dynamical processes in heavy-ion collisions~\cite{An:2023yfq}. The consequence of complexity due to fluctuations in effective potential shall also be explored in a finite-size system~\cite{An:2017brc}.  We leave these for future work.

\begin{acknowledgments}
\sect{Acknowledgments}
We thank G. Kovács, A. Moro, A. Sorensen and M. Stephanov for helpful conversations. We acknowledge the support and hospitality of the \emph{School of Mathematics and Statistics} of the University of Glasgow (X.A. and G.L.), the {\em Theoretical Physics Division} of National Centre for Nuclear Research (F.G. and G.L.), and the {\em Isaac Newton Institute for Mathematical Sciences} of the University of Cambridge (X.A., F.G., and G.L., supported by EPSRC grant EP/R014604/1). X.A. is supported by the European Research Council (ERC) under the European Union’s EU Horizon 2020 research and innovation program (Grant No.~101089093/project acronym: High-TheQ), and the National Science Centre (Poland) under Grant No.~2021/41/B/ST2/02909. G.L. is supported by INFN IS-MMNLP and IS-GAST. 
\end{acknowledgments}

\bibliographystyle{JHEP}

\bibliography{references}
 
\end{document}